\documentclass[12pt, preprint]{aastex}
\begin{document}
\title{The High Energy Budget Allocations  in Shocks and GRB }
\author{David Eichler\altaffilmark{1},
Dafne Guetta\altaffilmark{2}, Martin Pohl\altaffilmark{3,4}}
\altaffiltext{1}{Physics
Department, Ben-Gurion University, Be'er-Sheva 84105, Israel;} \altaffiltext{2}{Osservatorio astronomico di Roma, v. Frascati 33,
00040 Monte Porzio Catone, Italy.}
\altaffiltext{3}{Institut f\"ur Physik und Astronomie, Universit\"at
Potsdam, 14476 Potsdam-Golm, Germany}
\altaffiltext{4}{DESY, 15738 Zeuthen, Germany}

\begin{abstract}
The statistical distribution  of energies among particles responsible  for long Gamma Ray Burst (GRB) emission is
analyzed in light of recent results of the Fermi Observatory. The allsky flux, $F_{\gamma}$, recorded by the Gamma Ray Burst Monitor (GBM) is shown, despite its larger energy range, to be not significantly larger than that reported by the Burst and Transient Explorer (BATSE), suggesting a relatively small flux in the 3 - 30 MeV energy range.  The present-day energy input rate in $\gamma$-rays recorded by the GBM
from long GRB is found, assuming star-formation rates in the literature, to be  $\dot W(0)=0.5 F_{\gamma} H/c = 5 \times 10^{42}\ \rm{erg/Mpc^3 yr}$.  The Large Area Telescope (LAT) fluence, when observed, is  about 5-10\% per decade of the total, in good agreement with the predictions of saturated, non-linear shock acceleration.

The high-energy component of long GRBs, as measured by Fermi,  is  found to contain  only  $\sim 10^{-2.5}$ of the energy needed to produce ultrahigh-energy cosmic rays (UHECR) above 4 Eev, assuming the latter to be extragalactic,
when various numerical factors
are carefully included, if the cosmic ray source spectrum has a spectral index
of -2.  The observed $\gamma$-ray  fraction of the required UHECR energy is even smaller if
the source spectrum is softer than $E^{-2}$.

The AMANDA II limits rule out such  a GRB origin for UHECR if  much more than   $10^{-2}$ of the cosmic ray energy goes into
neutrinos that are within, and simultaneous with,
the $\gamma$-ray beam.
 It is suggested that "orphan" neutrinos out of the $\gamma$-ray beam might be identifiable via orphan afterglow { or other wide angle signatures of GRB in lieu of coincidence with prompt $\gamma$-rays}, and it is recommended that  feasible single neutrino trigger criteria  be established to search for such coincidences.
\end{abstract}
\keywords{gamma rays, cosmic rays}

\section{Introduction}

The high energy range of the cosmic ray (CR) spectrum is broken into three parts: The knee-to-ankle segment extends roughly from  $10^{16}$ to $4\times 10^{18} $eV, and the ultrahigh energy  CR (UHECR) above the ankle  can be further segmented into those below the GZK cutoff at about $4 \times 10^{19}$eV, and those above it. Those above the ankle  are believed   to be extragalactic, showing both a flattening of slope in the observed flux, and little anisotropy below the GZK cutoff, and some anisotropy towards the local supercluster above it. Both  UHECR components require particle acceleration well
beyond the energies at which $\gamma$-rays themselves are detected. The
cosmic rays (CR) beyond the ankle   have
 been attributed to active galactic nuclei,  but this is not the case for CR in the
range $10^{15} - 10^{18}$~eV, whose origin remains a mystery. In this energy range,  CR are
probably Galactic in origin and do not fill intergalactic space as the CR
above the ankle  do.

Gamma-ray bursts (GRB) have been considered by some authors (e.g. Levinson and Eichler 1993,   Milgrom \& Usov 1995, Waxman 1995, Vietri 1995) as sources of
UHECR.
Levinson and Eichler (1993), while not taking a position on whether GRBs could account for UHECRs above the ankle (despite giving this possibility serious consideration),  suggested tha suggested that those just below are due to Galactic GRB. {Doubts that GRB could supply the highest-energy cosmic rays include the total energetics, discussed in this paper, adiabatic losses, which lower the maximum energy should the acceleration be in a compact region, and Galactic isotropy,  which is discussed in a forthcoming paper.}  Later authors (Milgrom \& Usov 1995, Vietri 1995, Waxman 1995) focused on the possible connection between GRB and cosmic rays above the GZK cutoff.
GRBs have also been considered as sources of  UHE neutrinos (e.g. Eichler 1994,
Waxman \& Bahcall 1997,  Eichler and Levinson, 1999). Levinson and Eichler, The detection of UHECR and/or UHE neutrinos in association with GRBs would provide valuable information.

In recent years there have been new and improved data. In particular, the LAT detector on the Fermi observatory provides information about the  energy in non-thermal tails of particle distributions, and the Gamma Ray Burst Monitor (GBM), which can measure energies up to 30 MeV, provides a more reliable  measurement of the GRB bolometric luminosities. AMANDA II has been operational for over 1000 days and has set limits on the neutrino output of GRB.
This paper takes another look at GRB and UHECR energetics in the context of previous suggestions.
In the steady state, the total demands of local power per unit cosmic volume  on sources of UHECR  above the ankle, $\dot W_{\rm CR}$, can be written in terms of the measured allsky UHECR flux in any specified energy range, $F_{\rm CR}$, as
\begin{equation}
\dot W_{CR}
\equiv\left(\tau_{\rm CR}H\right)^{-1}\left(F_{\rm CR}H/c\right) C_B
=\left(\tau_{\rm CR}H \right)^{-1} \left(\frac{F_{\rm CR}}{F_{\rm GRB}}\right)\left(\frac{F_{\rm GRB}H/c}{\dot W_{\rm GRB}}\right)C_B{\dot W_{\rm GRB}}
\end{equation}
assuming the UHECR fill intergalactic space. Here $\tau_{\rm CR}$ is the cosmic-ray energy loss time due to interactions with the microwave background. The quantity
$C_B$ is the bolometric correction assuming the source produces a spectrum typical of familiar shock acceleration. The ratio $\frac{\dot W_{\rm CR}}{{\dot W_{\rm GRB}}}$ is thus given by a product of four separate dimensionless numbers.  Below, we determine the values of $\frac{F_{\rm CR}}{F_{\rm GRB}}$, $  \frac{F_{\rm GRB}H/c}{\dot W_{\rm GRB}(0)} $, $ \left(\tau_{\rm CR}H \right)$, and $ C_B$. We find that for cosmic rays above the ankle,  $\frac{F_{\rm CR}}{F_{\rm GRB}}$ is 4, $ \frac{F_{\rm GRB}H/c}{\dot W_{\rm GRB}(0)}$ is about 2 for popular estimates of cosmic star formation rates, $(\tau_{CR}H)^{-1}$ is between 5 and 8 for CR above the ankle, and more for CR above the GZK cutoff,  and $ C_B$ is between 4 and 20, depending on assumptions about the acceleration. Altogether, the value of $\frac{\dot W_{\rm CR}}{{\dot W_{\rm GRB}}}$ is of order $10^{2.5}$ to $10^3$,  despite the fact that each of the four factors may be modest.

\section{Particle Spectra in GRBs}

The average cosmic energy density in $\gamma$-rays from  GRB is a directly
measurable quantity, independent of distracting uncertainties in the beaming
angle, average energy, or rate of GRB. For a sample of hundreds of GRB or more,
and spectral coverage from 8 keV on upward, as we now have, we can measure the
total allsky flux in GRB merely by summing over all events {within} a given time
interval.
Most of the {photon energy in GRB detected by Fermi}
is in the GBM  range. We have considered all long ($T_{90} > 2$ seconds) bursts
detected by the  Fermi Gamma-Ray Burst Monitor from August 2008 until February
2010 (see Table 2 of Guetta et al. 2010).
The fluences have been collected from the literature (mainly GCN circulars) and
usually are given
in the 8 keV--1 MeV energy range. There are $\sim 20$ GRBs with fluences given
in the 50-300 keV energy range. For GRB 080916C, GRB 090902B and GRB 100116A the
fluences are given up to 10 MeV. There exists the possibility of some additional
unreported fluence outside the quoted range, but we presume the quoted range is
selected so that the unreported fluence is relatively small. We have summed the
reported GRB fluences  and found the sum to be $3.0\times  10^{-3}$~erg/cm$^2$. Short bursts, which we do not include here, contributed another several percent.
{ The GBM field of view (FOV) is roughly $\Omega_{\rm GBM}= 2\pi$
steradians, and we can thus write the total allsky} flux as

\begin{equation}
F_{\gamma} = 4 \times 10^{-3} \left(\frac{2\pi\ \rm sr}{\Omega_{\rm
GBM}}\right) \rm{erg/cm^2/yr.}
\label{el}
\end{equation}
This is  below the [10-1000] KeV flux for all GRB that would have been derived  from BATSE data by a factor of $\sim 1.25$ (M. Schmidt, private communication to V.S. Berezinsky 2002; D. Eichler and D. Guetta, in preparation). In addition to differences in instrumental sensitivity and calibration,  some discrepancy  may be expected due to year-to-year fluctuations in GRB flux, because it receives much of its contribution from the several  brightest GRB in any given year (see below).  Although we continue to use the GBM number, we note that the error in this estimate may be a factor of order 40 percent or so.

 The important conclusion is that the 3-30 MeV energy band, which was not covered by BATSE, does not appear to contain enough flux to significantly affect our conclusions. It appears to be less than the flux below 3 MeV. This is consistent with the steep
post-peak spectra -- i.e. large Band parameter $\beta$ -- that is seen for many of the bursts.
As most of the GRB { flux is in} relatively bright GRB,  we can be
reasonably confident that  no significant flux is hidden in dim,
nondetectable GRB - unless they form a separate, as yet unidentified, class of
events, rather than merely the low-luminosity extension of what has already been
identified. There remains, however,  the logical possibility that most GRB are dominated by huge GRB that are so infrequent that we do not yet have a fair sample of them.

Of the 205 long GRBs detected by the Gamma Ray Burst Monitor  before
February 2010, only $\sim$~12
long bursts were detected above 30 MeV by the LAT telescope. We have estimated the fluences above 100 MeV of 10 GRBs detected
by LAT\footnote{These are 090902B, 080916C, 090926A, 090323, 090328, 080825C,
090217, 091003, 090626 and 091031. We included neither GRB 081215A, as it has a
boresight angle of $86^\circ$ - hence out of FOV of LAT, nor GRB100116A, as we
didn't find its LAT fluence} to be between 5\% and 30\% of the total GBM
fluence.
{ The sum of the 100 MeV to 10 GeV fluences in these 10 LAT-detected GRB is
$1.34\times10^{-4}$~erg/cm$^2$,
which is about 13.5\% of the total GBM fluence for the same GRBs,
{ $1\times 10^{-3}$~erg/cm$^2$.}}

{ An additional 9 GRB were not detected by LAT despite being "LAT-eligible",}
i.e. within the LAT field of view and as bright as those that {\it were}
detected by LAT. Their total  GBM fluence was $4.5 \times  10^{-4}$ erg/cm$^2$,
and the upper limit to the LAT fluence for these GRB was about 5\% of the total.
Thus, the overall { emissivity} in the 100 MeV to 10 GeV range is apparently
below 13\% of that in the GBM band, which is consistent with the view that GRB
typically (though not necessarily always) emit most of their photon energy in a
{ quasi-}thermal MeV peak (e.g. Levinson and Eichler 2000, Meszaros and Rees
2000, Ryde 2004, Ryde and Pe'er 2009). The total detected LAT fluence over 1.5
years corresponds\footnote{the  LAT field of view is $\sim 2.5$
steradians,} to an allsky flux of

\begin{equation}
F_{\rm LAT}/2 =  (2.25 \pm 0.9) \times 10^{-4} \,\left(\frac{2.5\ \rm sr}{\Omega_{\rm
LAT}}\right)\  \rm{erg/cm^2/yr.}
\label{LAT}
\end{equation}
which is about 11\% $\pm$ 4.5\% of the {\it total} inferred all-sky GBM flux,
cf. equation \ref{el}.  The error of $\sim 40\% \sim 0.7/\sqrt3$ is estimated on
the basis of the fact that more than 70\% of the total is dominated by only 3
GRB, (090902B, 080916C, and 090926A). As we include two decades of energy in the
above flux, the inferred flux per decade is $F_{\rm LAT}/2$.

The Fermi Gamma Ray Observatory has a sufficiently large dynamic range of photon
energy that it is beginning to provide a basis for comparison with theory of how
energy in violent, non-thermal astrophysical events is distributed among
radiative particles. Theory predicts power-law
particle spectra whose index depends  on the compression ratio of the
shock. As the acceleration may be very efficient, the compression ratio  may be
affected by the particle acceleration itself, so that the problem can become
non-linear.  Non-linear theory of shock acceleration (Eichler 1979, Eichler
1984, Ellison and Eichler 1985) predicts that about half of the internal energy
of the shocked fluid should be found in the energetic tail and the other half in
thermal particles. The basic reason  is that, because the energetic
particle distribution must be injected from the thermal pool, some of the
compression (about half as it turns out)  must be saved for viscous heating of
the latter,\footnote{\noindent{ unless the phase velocity of the waves that couple the
energetic particles to the fluid is negligible, in which case the thermal
particles can get much less than half. Detailed calculations (Ellison and
Eichler 1985) show that the phase velocity must in this case be extremely
small, approximately two or more orders of magnitude below the shock velocity.}}
otherwise the injection of fresh  particles gets choked off. While the
energetic tail receives of order 50\% of the total, it is generally distributed
over many decades, so that there is only of order $1/2N_D\lesssim 10\%$ per decade or less. This
becomes significant in the interpretation of isolated energy ranges of
high-energy emission. The energy in UHECR is likely to be most of the total only
if the spectrum is extremely hard - much harder than the conventionally applied
models of shock acceleration - because the UHE range is so many decades above
the low-energy cutoff.

The  LAT fluence, when observed, appears to be only about $10-20\%$ of the total GBM fluence in the 100 MeV-10 GeV range,
or about $5-10\%$ per LAT decade, which is perhaps less than expected from
certain  early synchrotron emission models. The 100 MeV--10 GeV fluences observed by LAT may or may not be a good proxy for the total output of UHECR and/or UHE neutrinos. As opposed to the GBM flux,
which may represent a thermal pool, the LAT fluxes measure the non-thermal tail
of the energy distribution in the  GRB primary charged particles, which, if hadronic, is the part that could
contribute to the UHECR and UHE neutrino fluxes.  However, there still remain the uncertainties of the electron-to-proton ratio, the
microscopic details of the injection process, and the optical depth at which the
particle acceleration takes place. It is possible that degradation of
high-energy particles and/or photons also contributes to the thermal component. The point nevertheless remains
that maximally efficient shock acceleration over a large range of high energy is
generally limited from above by the large number of decades in the overall energy range of the acceleration process, to at most 5-10\% per decade of particle energy, so it is consistent with the  spectra of the LAT sources without the need to invoke pair opacity.  This necessitates a considerable bolometric correction, of order $1/2N_D\lesssim 10\%$, when comparing the total GRB flux with cosmic ray energies in a given energy decade.

 Were pair opacity  the only uncertainty, we might reasonably say that the LAT flux is a lower limit, and the GBM flux is  an upper limit, but this upper limit would include three decades in the GBM range, plus $N_p$, the number of decades in the range subject to pair opacity.  As the total GBM flux is only 20 times the LAT flux per decade, the upper limit and lower limits so obtained would be rather close together.  In our view, the main uncertainties in using LAT fluence as a proxy for UHECR output are the uncertainty in the electron-to-proton ratio and the possibility that only electrons contribute to the LAT $\gamma$-rays.

Relativistic shocks can accelerate charged particles, and, in the case of
efficient stochastic scattering a spectral index of -2.3 is generally predicted
(Kirk and Schneider 1987, Bednarz and Ostrowski  1998). Because the energy
spectrum is convergent at high energy, it need not affect the compression
ratio. It follows that only a small fraction $[\sim E_{\rm UHE}/E_{\rm
th}]^{-0.3}$ of {the relativistic} shock's energy budget can be apportioned to the highest
decade of the energy range.
  We  conclude that the measured LAT fluxes are consistent with maximally
efficient shock acceleration, if the GBM is interpreted as a thermal particle peak or as a catch-basin for pair cascade end products. They are also consistent with the $E^{-2.3}$
spectra expected from relativistic shocks, since they lie nearly 3 orders of
magnitude above the thermal peak and contain about 10\% of the total flux.

Another important energy reservoir for GRB is the kinetic energy of the blast, $E_{KE}$, which is the ultimate source of shock energy for producing UHECR.
As the proton-to-electron efficiency ratio $\epsilon_p/\epsilon_e$,
the proton to prompt-gamma-ray efficiency ratio  $\epsilon_p/\epsilon_{\gamma}$, and
the blast energy to prompt $\gamma$-ray  efficiency ratio  $\epsilon_{KE}/\epsilon_{\gamma}$
could each conceivably be large, no constraint can rigorously be placed  on the proton
energy budget from $\gamma$-rays, so the latter could come from inverse Comptonizing electrons.
Afterglow calorimetry { models} (Frail et al. 1997, Waxman et al. 1998,
Freedman and Waxman, 2001)  typically put  $\epsilon_{\rm KE}/\epsilon_{\gamma} \sim1$, but
see Eichler and Waxman (2005), and below. Also, Eichler and Jontof-Hutter (2005) argue
on the basis of correlation between $\epsilon_{\rm KE}/\epsilon_{\gamma}$
(as derived from afterglow calorimetry) and spectral peak, that
$\epsilon_{\rm KE}/\epsilon_{\gamma} $ is in fact nearly an order of magnitude
below previous estimates, and this is supported by the detailed analysis by Vergani et al. (2008) of GRB 060418. Simulations of relativistic shocks typically give
$\epsilon_p/\epsilon_e\sim 1$  (Spitkovsky 2008), but, as shocks responsible for
UHECR are more likely to be non-relativistic, this result may not apply. The possibility that $E_{\rm KE}$ is much larger than the $\gamma$-ray output $E_{\gamma}$, necessary to make the GRB-UHECR connection, is not consistent with many claims that $E_{\rm KE}\lesssim E_{\gamma}$, but we cannot rule it out, as there remains the possibility that the injection of electrons into the diffusive shock acceleration process is of low efficiency (Eichler and Waxman 2005).

In Figure 1, we adopt the star formation rate (SFR)
\begin{equation}
R(z)=R(0)
\frac{46\exp(3.4z)}{\exp(3.8z)+45}\, F(z,\Omega_M,\Omega_{\Lambda})
\end{equation}
where
\begin{equation}
F(z,\Omega_M,\Omega_{\Lambda})=
[\Omega_M(1+z)^3 +\Omega_{\Lambda}]^{1/2}/(1+z)^{3/2}
\end{equation}
with $\Omega_{\Lambda}=0.7$,  $\Omega_M=0.3$ (Porciani \& Madau 2001), and, for  comparison, that of Hopkins and Beacom (2006, 2008), which gives very similar results. Assuming the GRB energy input rate $\dot W(z)$
scales as  $R(z)$, we plot the expected energy contribution per unit time of GRB
to the present cosmic energy density $\dot W(z)/(1+z)$ as a function of redshift.
In the second frame, we calculate the cumulative contribution at redshift less than z,
\begin{equation}
\eta(z)\equiv\frac{\int^{z}_{0}[\dot W(z)/(1+z)][dt/dz]dz}{
\int^{z(T)}_{0}[\dot W(z)/(1+z)][dt/dz]dz}=\frac{\int^{z}_{0}[\dot W_{\gamma}(z)/(1+z)][dt/dz]dz}{
F_{\gamma}/c},
\end{equation}
where $T=13$~Gyr. Both SFR give a present day value for $-d\eta/dt$ of
\begin{equation}
-\dot\eta=(d\eta/dz)(-dz/dt) \approx 0.5 H \approx 0.04\ {\rm Gyr^{-1}}.
\end{equation}
The present day energy input from GRB  in the GBM range is thus given by
\begin{equation}
\dot W_{\gamma}(0) = -\dot\eta F_{\gamma}/c = 0.5F_{\gamma}H/c= 5 \times 10^{42} \rm{erg/Mpc^3 yr}
\end{equation}
which may be somewhat below some previous estimates. {Such a difference would appear if GRB with known afterglows are systematically brighter than those with unknown afterglow, for it is the former group that is the basis for estimating the typical isotropic equivalent energies  of GRB. This is connected to the claim  (Nakar and Piran, 2005, Band and Preece, 2005) that GRB with unknown redshift are less energetic on the average, even if placed at a very high redshift, than those of known redshift, which are reported to lie on the Amati relation.} On the other hand, Schmidt (2002, private communication to V.S. Berezinsky) obtains $ \dot W_{\gamma}(0)= 6 \times 10^{42} \rm{erg/Mpc^3 yr}$, in good agreement with our value.

We now compare the total flux in GRB radiation  to the flux per decade in ultrahigh
energy cosmic rays. The allsky differential number flux df(E)/dE in
UHECR is given by
\begin{equation}
E^3df(E)/dE = 4\pi \times 4 \times  10^{27}\,
\left(\frac{E}{10^{19}\ \rm eV}\right)^{0.3}\ \rm
eV^2/cm^2/yr
\end{equation}
(Abraham et al. 2010a),
in the range 4 to 40 EeV. This translates to {an allsky} energy flux
$F_{[4,40]}$ over the interval
$4\times 10^{18}\ {\rm eV}\le E \le 4\times 10^{19}\ {\rm eV}$) of

\begin{equation}
F_{[4,40]}=\int^{40EeV}_{4 EeV} dE\ E\,f(E) = 1.7 \times 10^{-2}\  \rm
ergs/cm^2/yr=4.0  \times F_{\gamma}.
\end{equation}

The horizon for UHECR in the 4-40 EeV range can be estimated by noting that the logarithmic energy-loss
rate,  $  \tau_{CR}^{-1}  + H $ in that range,  mostly due to pair production from CMB photons
{(e.g. Dermer and  Menon, 2009)}, is given by
\begin{equation}
\frac{\dot E}{E} = -(1+z)^3 (c/{\rm Gpc}) \frac{0.74 +1.8\, \ln(y/2)}{\sqrt{y}} +
\frac{d\ln(1+z)}{dt}
\sim  -4.2\,(1+z)^3\,H - H.
\end{equation}
 Here $H$ is the
Hubble constant, tentatively assumed to be constant at small z, $y\equiv E$~in EeV,
and we for convenience conservatively approximate the coefficient of loss due to pair
production, $[0.74 +1.8\,\ln(y/2)]/y^{1/2}$, as 1.0 at E$\ge 4$~EeV.  The second term on the right hand side represents loss due to cosmic expansion. Substituting $dln(1+z)/dt $ for H into equation (9) and solving shows that
an UHECR injected at E(z) at redshift (1+z) has a present day energy $E_o$ of
\begin{equation}
E_o=(1+z)^{-1}\,\exp\left(-1.4\,[(1+z)^3-1]\right)E(z).
\end{equation}
The horizon, defined to be the value
of z at which $(1+z)E_o/E(z)=1/e$, is thus at a redshift of $z_h=0.2$, and the energy
attenuation beyond this horizon is steep.

Assuming the source spectrum $S(E,z)$ to be constant in time, to be multiplied by the star formation rate R(z),    the total number per comoving volume of cosmic rays at the present time $N(E_o)$, is given by

\begin{eqnarray}
N(E_o)dE_o
&=\int_0^{\infty}S[E(z)]\,dE(z)\,R(z)\,dt\\
&=S(E_o)\,R(0)\,\int_0^{\infty}\frac{S[E(z)]\,dE(z)\,R(z)}{S[E_o]\,R(0)\,(1+z)\,H}\,dz
\end{eqnarray}
Assuming $S(E)\propto E^{-\alpha}$, then
\begin{equation}
\frac{S[E(z)]dE(z)}{S(E_o)dE_o}
=(1+z)^{-\alpha+1}\exp{\left(1.4({1-\alpha})[(1+z)^3-1]\right)}
\end{equation}
The  present number density  of CR at energy $E_o$ is therefore given by
\begin{equation}
N(E_o)
=S(E_o)R(0)\int_0^{\infty}\frac{(1+z)^{-\alpha}}{H}\,\exp{\left(1.4({1-\alpha})[(1+z)^3-1]\right)}\frac{R(z)}
{R(0)}dz
\end{equation}

Evaluating the integral numerically for $\alpha=2.4$,
 we find
\begin{equation}
S(E_o)R(0)
=6.45N(E_o)H
\end{equation}
and similarly $ \dot W(0)\approx 6.45 FH/c \approx 2.5\times 10^{44}{\rm erg/Mpc^3 yr}$. Uncertainty in $\alpha$, $2.0\le \alpha \le 2.7$, introduces about a 20\% uncertainty in this estimate. However, the observed CR spectral index between 4 and 40 EeV, -2.7, suggests that the slope cannot be much flatter, because the pair creation loss time is nearly energy independent.


Above $4\times 10^{19}$ eV, in the trans-GZK regime, the flux, to within the Auger
error bars, is
\begin{equation}
E^3\,df(E)/dE= 4\pi \times 5 \times 10^{27}\
\left(\frac{E}{3\times 10^{19}\ \rm eV}\right)^{-1.2}\ \rm eV^2/cm^2/yr.
\end{equation}
This implies an energy flux
\begin{equation}
F_{>\rm GZK} = 1.1 \times 10^{-3}\ \rm ergs/cm^2/yr.
\end{equation}
 This flux  may be supplied only by
recent $(z\ll 1)$ GRB, if any, because the sources must come from within $c\,t_{\rm GZK}
\sim 200$~Mpc or so (e.g. Olinto et al. 2010), so the contributing GRB must therefore
have occurred  within the past $6.18\times 10^8$~yr, i.e. at $z\lesssim 0.05$. The horizon length above 40 EeV is very sensitive to energy, and therefore to the exact energy calibration of the Auger detector. For the purposes of a rough estimate, we write the average horizon length as $(200\ {\rm Mpc})\,l_{200}$. Using the above
SFR, we roughly estimate the fraction of this recent contribution to the total to be $\eta (z[t_{\rm GZK}]) \approx t_{\rm GZK}\dot\eta  =0.025\,l_{200}$  and that the present rate of trans-GZK CR production is $F_{>\rm GZK}/c \eta (0.05) = 4.4 \times 10^{-2}/l_{200}c\ \rm ergs/cm^2/yr = 1.3\, l_{200}^{-1}\times 10^{44}\rm erg/Mpc^3yr$. This can be compared to other estimates, which range from $0.45(\alpha - 1)\times 10^{44}\rm ergs/Mpc^3/yr$ per unit interval in ln E above 30 EeV (Katz, Budnick, and Waxman 2009) to $1.67\times 10^{44}\rm ergs/Mpc^3/yr$ (Berezinsky 2008), if the energy calibration is set to optimize the fit to  the observed GZK cutoff. The uncertainty lies in the instrumental energy calibration and in the location of the high energy cutoff.

If most of the trans-GZK CR are heavy nuclei, as suggested by recent
analysis of AUGER data, (Abraham et al., 2010b), then their production
is limited to an even smaller lookback time (Stecker and Salamon, 1999), and the
  UHECR production demanded of individual GRB would be raised accordingly.
This issue is somewhat controversial at present, and we have conservatively taken the
UHECR to be protons for the sake of this discussion.

Finally, we  turn to the bolometric correction $C_B$.
The forward shocks of
relativistic blast waves that produce GRB afterglow, if they have a spectral
index of -2.3 from some minimum energy $E_{\rm th}$ through some ultrahigh energy $E_{\rm UHE}$,  would require a bolometric correction of $C_B=[E_{\rm UHE}/E_{\rm th}]^{0.3}$. If we take $log_{10}[E_{\rm UHE}/E_{\rm th}]$ to be 5, as in Dermer (2010) we would get $C_B\approx10$. (Dermer takes a relativistic spectral index of 2.2, in which case $C_B\approx 5$, illustrating how sensitive $C_B$ is on spectral index.)
 The less energy-demanding source of the UHE primaries is generally  non-relativistic
shocks, which can have flatter spectra (or possibly radiative or cosmic-ray-lossy shocks, which don't conserve energy from upstream to downstream). These
could, for example, be internal shocks in a GRB baryonic outflow
 (Levinson and Eichler, 1993,  { Rees \& Meszaros 1994}), or the subrelativistic end phase of GRB
blast waves. Here the minimum energy is at most $\sim E_{max}/\Gamma_S m_pc^2$, where $\Gamma_S$ is the shock Lorentz factor, and the bolometric correction or any given decade of energy, e.g. [4,40]EeV, is about $[\log_{10}E_{max}-\log_{10}E_{\rm
th}] \sim [11-\log_{10}(\Gamma_S)]\sim 8$.

We also consider the energy per decade inferred from  the Fermi LAT detections
in the energy decades 100 MeV to 10 GeV as the basis for a representative
estimate of the amount of the energy per decade that could be expected for UHECR
and UHE neutrinos at even higher energies. The GBM flux, which spans several
orders of magnitude of photon energy and  which is an order of magnitude higher
than the LAT flux, may tentatively be interpreted as a measure of the total
energetics of GRBs, as it can be associated with thermal photons  and/or pair
cascade end products that emerge from a photosphere. For example, the GBM flux, which appears to have a quasithermal spectrum peaking around 1 MeV, may be the catch-basin of all of the radiation that was converted into pairs. For GRB unseen by LAT, this can be as much as all the fireball energy. In cases where the GRB is detected by LAT, the observed GBM-to-LAT
flux ratio is consistent with
the assumption that shock acceleration could put at most of order $1/2N_D$ of its
energy budget  at any given decade well beyond the thermal peak of
cosmic-ray energy. We tentatively make the conventional best-case assumption for
shock acceleration [tentatively leaving aside the extreme case of negligible phase velocity,
(Ellison and Eichler, 1985)] that the spectral index is -2.0, which represents a
spectral distribution of equal energy per decade of individual particle energy.

Altogether, assuming $C_B =8$, and  combining it with  $\frac{F_{\rm UHECR}}{F_{\rm GRB}}=4$, $ \frac{F_{\rm GRB}H/c}{\dot W_{\rm GRB}(0)}=2$, and $(\tau_{\rm CR}H)^{-1}\sim 6.5$, we would obtain $\frac{\dot W_{\rm UHE}(0) }{\dot W_{\rm GRB}(0)}=416 $.  If we replace $C_B\sim 8$ with the correction that would obtain using the LAT proxy for energy per decade, $F_{\rm GBM}/[F_{\rm LAT}/2]\sim 20$,  we would obtain $\frac{\dot W_{\rm UHE}(0) }{\dot W_{\rm GRB}(0)}=1000$.

It may be possible to circumvent much of the bolometric correction by assuming that only the highest energy cosmic rays escape the shock at the acceleration site, while the rest are recycled by adiabatic expansion. This might give a source spectrum that is much harder than the post-shock spectrum.  However, this does not seem to be the case for supernova remnants (SNR), if they are the source of low energy CR, as the Galactic energy budget in CR is only about 3\% of the energy budget of SNR. Rather, it would seem that adiabatic losses of UHECR (which we have not included) would exacerbate the bolometric correction. There are ways in the literature to get a very hard spectrum\footnote{Although there has been no observational evidence for this, it is
mathematically possible for shocks to dissipate most of their energy into cosmic
rays of the highest energy that escape the system (Eichler, 1983, Ellison and
Eichler, 1985). The escaping CR have the compression-enhancing effect of
radiative losses, and the spectrum is much harder than $E^{-2}$. This does not,
however, lessen the energy requirement above 4 EeV, which still challenges the hypothesis of a GRB origin for UHECR.
Other "ultrahard" acceleration  scenarios exist for baryonic acceleration in GRB which
could produce much harder spectra than shock acceleration. In the neutron
pick-up scenario ( Eichler and Levinson, 1999, Levinson and Eichler, 2003), neutrons that leak out of the
surrounding matter into the fireball are picked up by a high-$\Gamma$ Poynting
flux when a charged particle within the high-$\Gamma$ flow collides with them.
After half a gyration { or isotropization}
in the high-$\Gamma$ flow they end up with $\sim
\Gamma^2\,mc^2$ of energy. A collisional avalanche ensues in which each collision
between a charged particle and a neutron {produces} further charged particles at
$\sim \Gamma^2mc^2$. Most of the dissipation of
the flow is {through} high-energy hadronic collisions, and
much of the energy loss is
in the form of UHE neutrinos.
Pick-up of ex-neutrons (and/or other ex-neutrals) can accelerate hadrons by a factor of $\Gamma^2$  as long as it does not exceed $2\sigma^{2/3}$ ( Eichler 2003), where $
\sigma\, mc^2/Ze$  is the maximum potential drop across the MHD flow (Michel,
1983).  For GRB central engines with magnetic fields in excess of
$\sigma \ge 10^{15}$, ultrahigh energies fall within this limit. The maximum Lorentz factor of GRB fireballs is still unknown.
Repeated charge conversion can lead to multiple encounters  of this sort and accelerate particles up to the limit, potentially leading to a hard spectrum (Derishev et al. 2003). Similarly, multiple encounters with independent cells in relativistic turbulence of Lorentz factor $\gamma$ can achieve similar repetition of amplification by $\gamma^2$ (Dermer, private communication, 2010).} and this is an interesting issue. But the inferred value for $\frac{\dot W_{UHE}(0) }{\dot W_{GRB}(0)}$ is quite large even without any bolometric correction.

\section{UHE Neutrinos}

We now briefly consider UHE neutrino detectability from GRB.  Muon neutrino
collection for IceCube is most efficient per unit energy in the range $1\le
E_{\nu} \le 100$ erg  (Gaisser, 2009), where the effective
collecting area is  about 0.3 to $\sim 10^2$ m$^2$. In what follows, we consider the
contribution in these two decades, assuming that the muon-neutrino cross section
is  $1(E/\rm erg) m^2$ in this region, which is accurately averaged over the
energy range $1\le E_{\nu} \le 100$ erg, and to within a factor of 2 at any
given energy within. The minimum flux to produce 1 count per year is therefore
$(1-2) \times 10^{-4}$ erg/cm$^2$/yr.
 At high energies, the expected ratio of photon energy to
neutrino (plus anti-neutrino) energy is about 4:5, because neutral pions
are about half as numerous as charged ones {and put all their energy into 2
photons whereas the charged pions put only about half their energy into a muon
neutrino/antineutrino pair, and divide the other half about equally between two neutrinos and a charged lepton, the latter of which probably radiates its energy into photons.} Assuming oscillations thoroughly mix the neutrino energy among the three neutrino types, the ratio of muon neutrino energy to photon energy is then about 5:12.
The measured LAT flux, if from hadrons or their collision products, combined with the hypothesis of a
flat energetic hadron spectrum,  would suggest a flux of
$ \sim 2[\Omega_{\rm IceCube}/4\pi][\Omega_{\rm LAT}/(2.5\,{\rm sr})]^{-1}   $
counts per year.
Here
 $\Omega_{\rm IceCube}/4\pi$, the fraction of sky available for neutrino
detection, may be close to unity in the context of GRB, where there is so small
a noise problem from atmospheric muons.

The number of neutrino events per year allowed in
IceCube from GRB is limited
by the lack of neutrino events in AMANDA correlating with GRB.  AMANDA II has had $\sim
3.8 $ years of live observing time from 2000 through 2007, and has an effective
collection area of about 1/20 of IceCube 80 (De Young, et al., 2008).
 { Despite the differences in
the spectral response of IceCube and AMANDA, and the varying availability of the contemporaneous
GRB monitoring needed
to eliminate blind searches, the failure to
identify a single neutrino event {in AMANDA } associated with known GRB (Achterberg et al. 2008)}
 limits (to modest confidence) the expected GRB contribution to the IceCube
detection rate to $\sim 5$ events per year, unless the neutrinos are a) out of the $\gamma$-ray beam (Eichler and Levinson, 1999) or b) dominated by large episodic events more than a year or two apart. [A $(1-e^{-n})$ confidence limit would be n times higher.]

This limit corresponds to a UHE
neutrino flux per energy decade of less than $10^{-3}$ erg/cm$^2$/yr.  This is less than $5\times 10^{-3}$ of the cumulative energy density, $F_{[4,40]}/c\eta(0.2)$, that must have been put into UHECRs in the 4-40 EeV range to maintain the observed $F_{[4,40]}$,  and about $2\times 10^{-2}$ of that needed to maintain the observed trans-GZK flux, $F_{>GZK}/c\eta(z_{GZK})$.  This would then rule out models in which GRBs produce the UHECRs in region in regions of
moderate to high optical depth to hadronic interactions and subsequent photons, unless the UHECR energy spectrum is harder than $E^{-2}$ in the  $10^{15} - 10^{20}$ eV range.

Note that this limit does { not} apply to "orphan" neutrinos from   obscured  or otherwise unseen GRB (Eichler and Levinson 1999) as there is no associated GRB to which to "pin" the neutrino burst. Collimation of fireballs within the envelope of  host stars, now believed to almost certainly take place,  would naturally lead to wider neutrino beams than the $\gamma$-ray beams if there is proton acceleration at collimation shocks. Similarly, shocks in an accompanying baryonic outflow (Levinson and Eichler 1993) may produce  a wide neutrino beam while accompanying $\gamma$-rays are stopped. There may  eventually be "$\gamma$-quiet" ways to confirm that a UHE neutrino is associated with a GRB  (Eichler and Levinson 1999, Guetta and Eichler 2010). For example, an orphan photon afterglow is detectable within several days, and, at $E_{\nu}\ge$ TeV, the atmospheric background within the IceCube angular resolution, less than 1 per square degree per year, is unlikely to cause confusion.

{ As noted early on (Eichler, 1994), {\it individual GRB} are unlikely to produce more than one neutrino count unless they are exceptionally energetic to the extent of being rare, and this is important for {\it prospective } identification of HE muon/neutrino events with GRB. The brightest { individual} LAT fluence was $4.3\times 10^{-5}$erg/cm$^2$,\footnote{GRB 090926A} which would correspond to an expected neutrino fluence of  about 0.2 by the above estimation procedure, and this is consistent with the assumption that several percent of GRB-related neutrinos would arrive paired, which is important for follow-up searches for paired neutrinos (Franckowiak et al, 2009).  However,the 0.1 to 0.2 probability applies only to the four brightest examples. The average for the 19 LAT-eligible long GRB, on the other hand, was limited from above by $8\times 10^{-6}$erg/cm$^2$, which would imply a pairing fraction less than 1/19. The number of expected neutrino pairs from GRB, if LAT fluences are used as a basis for estimate, is thus less than 1/yr, though obviously uncertain.}
{ Because neutrino pairs from GRB are (under reasonable assumptions) probably rare, as discussed above, we suggest therefore that single HE and UHE neutrino counts in IceCube be subjected to  follow up searches for "sibling" orphan  afterglows- for example, with robotic telescopes  and/or high energy cuts. The selection criteria for choosing a tractable set of single events to be followed up  is a non-trivial matter, but a  high energy threshold seems to be one obvious approach.}

The Dec. 27 2004 flare from the soft $\gamma$-ray repeater SGR 1806-20, which, at a fluence
of about 1 erg/cm$^2$, produced the equivalent of 250 years of GRB prompt-emission
fluence, did not register in AMANDA II, and the limit this sets on
neutrino fluence to $\gamma$-ray fluence ratio is $3\times 10^{-3}$ (Achterberg  et al. 2006).
Present-day understanding of GRB, which may (or may not) be sufficiently
similar to giant SGR outbursts to make the analogy, is thus compatible with
(but does not imply) a signal of less than $10^{-1}$ UHE neutrino events in
IceCube per year  from GRBs.

\section{Summary and Discussion}

We find that the prompt emission and total afterglow emission from GRB produce a
cosmic energy density that is two to three orders of magnitude below that
ascribed to processes that yield  ultrahigh-energy cosmic rays. This estimate
assumes that UHECR are generated with an $E^{-2}$
spectrum. For even a slightly steeper spectrum, the difference between the implied UHECR energy input  and the GRB photon energy input is even larger. It applies to the net production of UHECR after adiabatic losses.
Such losses are to some extent inevitable for shock acceleration, which relies
on compression of trapped particles, but the degree to which they affect the UHE
output has not, to our knowledge, been calculated.

The total LAT flux from GRB, if {it is of hadronic origin and
if GRB generate, in UHE neutrinos in the 1 to 100 TeV
range, the same energy per decade as what LAT detects in high energy $\gamma$-rays, implies a neutrino rate of about 2 events per year in IceCube.} Based on
the fact that the overall GBM flux is mostly from the brightest GRB, there is a
good chance that these neutrinos would mostly correlate with the small minority
of bright GRB rather than one of the much larger number of dim GRB.

Fermi can directly observe $\gamma$-rays from GRB up to about 100 GeV
beyond which an undetected additional component from, e.g., inverse Compton
scattering, could hide. Direct limits on TeV band emission (Aharonian et al. 2009)
from GRB are not
constraining on account of severe absorption by pair production
{in intergalactic space}, but the pair production
will feed an electromagnetic cascade that transports the radiation energy into
the
10--100 GeV band, where it would appear as extragalactic background. Recent
Fermi
measurements of that background indicate that the intensity in the 10-100 GeV
band
is about $5\cdot 10^{-10}\ {\rm erg/cm^2/s/sr}$ (Abdo et al. 2010), or an
allsky flux of
$2\cdot 10^{-1}\ {\rm erg/cm^2/yr}$.  This can be compared to the implied
full-sky flux of $\sim   10^{-1}\ {\rm erg/cm^2/yr}$ that is implied by
equation (4)  when  the entire energy in a E$^{-2}$ spectrum from 100 GeV to trans-GZK energies
is cascaded into the 10--100 GeV band.
This means that, whatever the origin of UHECR,
a significant fraction of them {\it may} have undergone high energy collisions.
This admits, for {\it non-GRB} scenarios of UHECR origin, the "thick target" scenario
- that much or most of the UHE particle acceleration in the universe is buried
at large optical depths, and that the  UHE electromagnetic radiation   we
observe, even in a particular episodic event, is just the part that stands out
at modest optical depth. The diffuse 10-100 GeV $\gamma$-ray background would, by itself,  admit much larger UHE neutrino fluxes, up to $\sim 250$ {IceCube events}
per year, even if the associated photons escape the acceleration site.

Although we have attempted careful estimates of what is known, the overwhelming
uncertainty in the above discussion is the spectral index of the cosmic rays
which we have taken as $E^{-2}$ for reference. The UHECR and neutrino outputs,
whose relevant energies are much higher than {those of} the $\gamma$-rays observed so
far, are so sensitive to the spectral index as to render relatively minor the
other considerations that we can discuss with some degree of knowledge.

The other chief uncertainty  is the optical depth at which hadronic particle
acceleration takes place. If most of it is at high optical depth, the
UHE-neutrino signal should be relatively strong. However,
the AMANDA limits on GRB-associated UHE neutrinos are relevant for any optical depth.

The main conclusions to be drawn are:

a) Identification of ultrahigh-energy cosmic rays in association with GRB would mean that GRB put more than $\sim 50$ times as much energy into UHE cosmic
rays above the ankle in the UHECR spectrum as into all $\gamma$-rays. If the spectral index {of accelerated particles}
is -2, then they put about
$10^{2.5}$ to $10^{3}$ times as much energy into cosmic rays over the full spectral range.
In view of the large energy requirements for the latter possibility, such a
result, should it come to pass, could be interpreted as evidence for ultrahard UHE hadronic acceleration
scenarios within the GRB, { as well as a huge baryonic blast wave component.}

b) The hypothesis of a GRB origin for UHECR by processes occurring over a range of optical depths that includes
moderate to high optical depth for the UHECR would need to confront the non-detection by
AMANDA II of any GRB-associated neutrinos.

{ c) Prospective searches for supernovae or GRB signatures with single neutrino events in IceCube are motivated by the chance that doublet events will be too rare.}

We are grateful to C. Dermer, F. Halzen, Y. Lyubarsky, A. Pe'er, B. Katz, and E. Waxman for helpful discussions. This research was supported by the Israel-US Binational Science
Foundation, the Israeli Academy of Science, and The Joan and Robert
Arnow Chair of Theoretical Astrophysics.

\begin{figure}
\includegraphics[width=0.7\columnwidth, keepaspectratio]{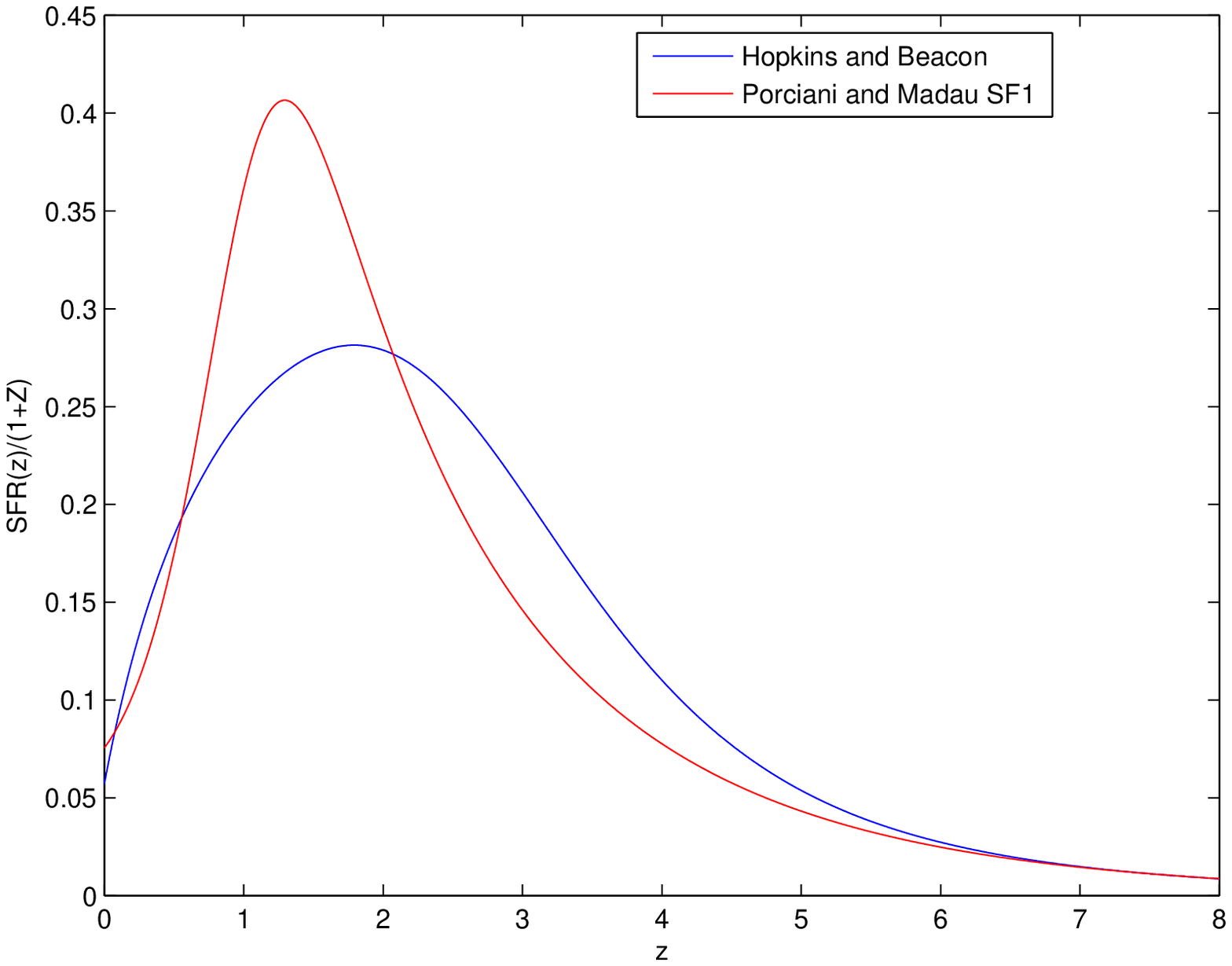}
\includegraphics[width=0.7\columnwidth, keepaspectratio]{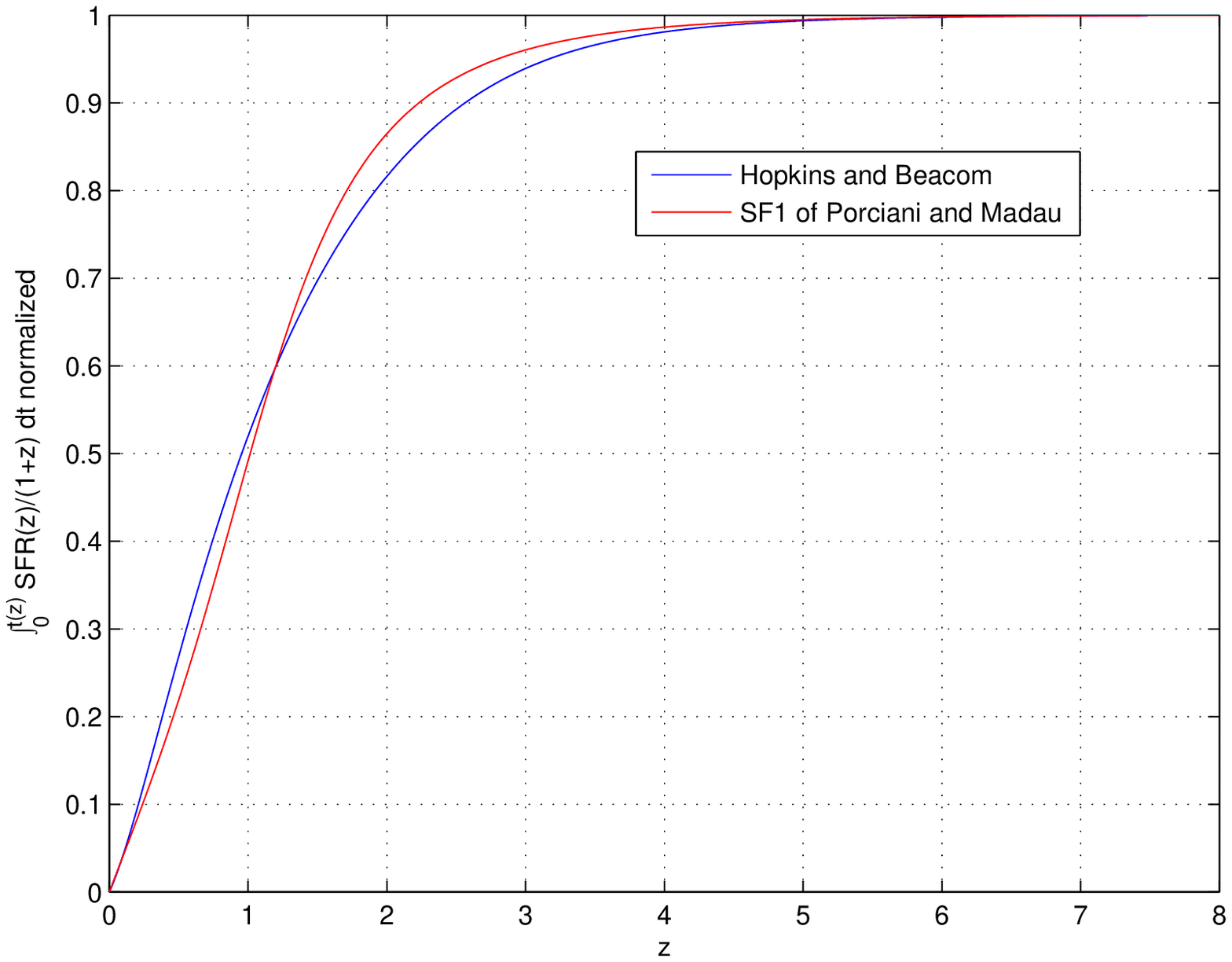}
\figcaption[FileName]{\label{f01}\footnotesize Upper panel:  Differential
contribution to the overall GRB fluence as a function  of the redshift. Lower
Panel: Integrated contribution to the overall GRB fluence as a function  of the
redshift. The assumed star formation rates are discussed in the text. }
\end{figure}

\section{References}

\noindent  Abdo, A. A. et al. 2010, Phys. Rev. Lett. 104, 101101

\noindent Abraham, J., et al., Pierre Auger Collaboration, 2010a, Phys. Lett. B, 685, 239; (arXiv 10.02.1975)

\noindent Abraham, J., et al., Pierre Auger Collaboration, 2010b, Phys. Rev. Lett. 104, 091191

\noindent  Achterberg, A., et al. 2006, Phys. Rev. Lett. 97, 221101

 \noindent  Achterberg, A., et al. 2008, ApJ 674, 397

\noindent Aharonian, F. 2009 A\&A 495, 505

\noindent Berezinsky, V.S. 2008, Advances in Space Research, 41, 2071

\noindent Bednarz J., Ostrowski M. 1998, Phys. Rev. Lett. 80, 3911

\noindent Band, D.L., Preece, R.D., 2005, ApJ 627, 319

\noindent Derishev, E. V.; Aharonian, F. A., Kocharovsky, V. V., Kocharovsky, Vl. V. 2003, Phys. Rev. D, 68, 043003

\noindent Dermer, C. 2010, astro-ph/1003.5318

 \noindent Dermer,  \& Menon 2009, "Radiation from Black Holes: Gamma Rays, Cosmic rays and Neutrinos" (Princeton Univeristy Press)

\noindent De Young, T., for the IceCube Collaboration. 2008, J. Phys. Conf. Ser., 136, 022046, arXiv:0810.4513v1 [astro-ph]

\noindent Eichler, D. 1979, ApJ 229, 419

\noindent Eichler, D. 1983, ApJ, 272, 48

\noindent Eichler, D. 1984, ApJ, 277, 429E

\noindent Eichler, D. 1994, ApJS 90, 877.

 \noindent Eichler, D. and Levinson, A. 1999, ApJ., L521, 117

\noindent Eichler,  D. arXiv:astro-ph/0303474

 \noindent  Eichler, D. \& Jontof-Hutter, D. 2005, ApJ 635, 1182

 \noindent Eichler, D. \& Waxman, E. 2005, ApJ, 627, 861

 \noindent Ellison, D. C. \& Eichler, D. 1985, Phys. Rev.Lett. 55, 273.

 \noindent Franckowiak, A. Akerlof, C., Cowen, D.F. Kowalski, M., Lehmann, R. Schmidt, T., Yuan, N., 2009, arXiv: 09090631

\noindent Frail, D. A., Kulkarni, S. R., Nicastro, L., Feroci, M., \& Taylor, G. B. 1997, Nature, 389, 261

\noindent Freedman, D. \& Waxman, E. 2001, ApJ 547, 922.

\noindent Gaisser, T. K. 2010, EP\&S, 62, 195

 \noindent Guetta,D. \& Eichler, D., 2010, ApJ 710, 392

 \noindent Guetta, D., Pian, E. \& Waxman, E. 2010 astro-ph/1003.0566

 \noindent Hopkins, A. M. \& Beacom, J. F. , 2006, ApJ 651, 142, erratum 2008 ApJ 682, 1486

 \noindent Kirk, J.G. and Schneider, P. 1987, ApJ, 315, 425

 \noindent Katz, B., Budnick, R., and Waxman, E. arXiv:0811.3759

 \noindent Levinson A. \& Eichler, 1993, ApJ 418, 386

 \noindent Levinson, A. \& Eichler, D. 2000, Phys. Rev. Lett. 85, 236.

 \noindent Levinson, A. \& Eichler, D. 2003, ApJ, 594L, 19L.

 \noindent   Meszaros, P. \& Rees, M. 2000, ApJ 530, 292

  \noindent Michel, F. C. 1983, Bulletin of the American Astronomical Society, Vol. 15, p.937

  \noindent Milgrom, M. \& Usov, V. V. 1995, ApJ 449, L37

    \noindent Nakar, E., \& Piran, T.  2005, MNRAS, 360, L73; 2004, arXiv:astro-ph/0412232

  \noindent Olinto, A. V., et al. 2010, Astro 2010: The Astronomy and Astrophysics Decadal Survey Science White Paper, no. 22; 2009 astro-ph/0903.0205

     \noindent Porciani, C. \& Madau, P. 2001, ApJ 548, 522.

   \noindent Rees, M. \& Meszaros, P.  1994, ApJ 430, L93

  \noindent  Ryde, F. 2004 ApJ 614, 827

  \noindent Ryde, F. \& Peer, A. 2009, ApJ 702, 1211

  \noindent Spitkovsky, A. 2008, ApJ 682, L5

  \noindent Stecker, F. \& Salamon, M. H. 1999, ApJ 512, 521

  \noindent Vergani, S.D., Malesani, D. Molinari, E. 2008, International Journal of Modern Physics, 17, 1343

  \noindent Vietri, M. 1995, ApJ 453, 883.

 \noindent Waxman, E. 1995, Ph. Rev. Lett. 75, 386.

 \noindent Waxman, E. \& Bahcall, J. 1997, Ph. Rev.Lett. 78, 2292.

  \noindent Waxman, E. Kulkarni, S. R. \& Frail, D. A. 1998, ApJ 497, 288


\end{document}